\documentstyle[fleqn]{article}

\title{Stability and noise in
biochemical switches}

\author{William Bialek \\
NEC Research Institute\\ 
4 Independence Way\\
Princeton, New Jersey 08540\\
{\it bialek@research.nj.nec.com}
}

\begin{document}

\maketitle

\begin{abstract}
Many processes in biology, from the regulation of gene expression in
bacteria to memory in the brain, involve switches constructed from
networks of biochemical reactions. Crucial molecules are present in small
numbers, raising questions about noise and stability. Analysis of noise
in simple reaction schemes indicates that switches stable for years and
switchable in milliseconds can be built from fewer than one hundred
molecules. Prospects for direct tests of this prediction, as well as
implications, are discussed.\end{abstract}

\section{Introduction} 
The problem of building a reliable switch
arises  in several different biological
contexts.  The classical example is the switching
on and off of gene expression during development
\cite{slack,lawrence}, or in simpler systems such as  phage
$\lambda$
\cite{ptashne,lambda-rev}.  It is likely  that the cell cycle should also
be viewed as a sequence of switching events among discrete states, rather
than as a continuously running clock \cite{cycle}. The stable switching
of a specific class of kinase molecules between active and inactive
states is believed to play a role in synaptic plasticity, and by
implication in the maintenance of stored memories
\cite{kennedy-orig,kennedy-rev}.  Although many details of mechanism
remain to be discovered, these systems seem to have several common
features.  First, the stable states of the switches are dissipative, so
that they reflect a balance among competing biochemical reactions. 
Second, the total number of molecules involved in the construction of the
switch is not large.  Finally, the switch, once flipped, must be stable
for a time long compared to the switching time, perhaps---for development
and for memory---even for a time  comparable to the life of the
organism.  Intuitively we might expect that systems with small numbers of
molecules would be subject to noise and instability
\cite{schrodinger}, and while this is true we shall see that
extremely stable biochemical switches can in fact be built from a  few
tens of molecules.   This has interesting implications for how
we think about several  cellular processes, and 
should be testable directly.

Many biological molecules can exist in multiple states, and biochemical
switches use this molecular multistability so that the state
of the switch can be `read out' by sampling the states (or enzymatic
activities) of individual molecules.  Nonetheless, these biochemical
switches are based on a network of reactions, with stable states that
are collective properties of the network dynamics and not of any
individual molecule.  Most previous work on the properties of
biochemical reaction networks has involved detailed simulation of
particular kinetic schemes
\cite{arkin98review,recent-schemes}, for example in discussing
the kinase switch that is involved in synaptic plasticity
\cite{lisman1}.  Even the problem of noise has been
discussed heuristically in this context \cite{lisman-goldring}.  The goal
in the present analysis is to separate the problem of noise and stability
from other issues, and to see if it is possible to make some general
statements about the limits to stability in switches built from a small
number of molecules.  This effort should be seen as being in the same
spirit as recent work on  bacterial chemotaxis, where
the goal was to understand how certain features of the computations
involved in signal processing can emerge robustly from the network of
biochemical reactions, independent of kinetic details
\cite{barkai-leibler}.

\section{Stochastic kinetic equations}

Imagine that we write down the kinetic equations for some set of
biochemical reactions which describe the putative switch.  Now let us
assume that most of the reactions are fast, so that there is a single
molecular species whose concentration varies more slowly than all the
others.  Then the dynamics of the switch essentially are one dimensional,
and this simplification allows a complete discussion using standard
analytical methods.  In particular, in this limit there are general
bounds on the stability of switches, and these bounds are independent
of (incompletely known) details in the biochemical kinetics.  It should
be  possible to make progress on multidimensional versions of the
problem, but the point here is to show that there exists a limit in which
stable switches can be built from small numbers of molecules.

Let the number of
molecules of the `slow species' be $n$.   All the different reactions can
be broken into two classes:  the synthesis of the slow species at a rate
$f(n)$ molecules per second, and its degradation at a rate $g(n)$
molecules per second; the dependencies on $n$ can be complicated because
they include the effects of all other species in the system.  Then, if we
could neglect fluctuations, we would write the
effective kinetic equation
\begin{equation}
{{dn}\over{dt}} = f(n) - g(n).
\label{kinetics}
\end{equation}
If the system is to function as a switch, then the stationarity condition
$f(n)=g(n)$ must have multiple solutions with appropriate local stability
properties.

The fact that molecules are discrete units means that we need to give
the chemical kinetic Eq. (1) another interpretation.   It is the mean field
approximation to a stochastic process in which there  is a probability
per unit time $f(n)$ of making the transition $n\rightarrow n+1$, and a
probability per unit time $g(n)$ of the opposite transition $n\rightarrow
n-1$.   Thus if we consider the probability $P(n,t)$ for there being $n$
molecules at time $t$, this distribution obeys the evolution (or
`master') equation
\begin{equation}
{{\partial P(n,t)}\over{\partial t}}
= f(n-1) P(n-1,t) + g(n+1) P(n+1, t) -[f(n) + g(n)] P(n,t) ,
\label{master}
\end{equation}
with obvious corrections for $n=0, 1$.
We are interested in the effects of stochasticity for $n$ not
too small. Then $1$ is small compared with typical values of $n$, and we
can approximate $P(n,t)$ as being a smooth function of
$n$.  We can expand Eq. (\ref{master}) in derivatives of the distribution,
and keep the leading terms:
\begin{equation}
{{\partial P(n,t)}\over{\partial t}}  =
{\partial\over{\partial n}}
\left\{
[g(n) -f(n)]P(n,t)
+{1\over 2} {\partial\over{\partial n} }[f(n) +
g(n)]P(n,t)
\right\} .
\label{diffuseq}
\end{equation}
This is analogous to the diffusion equation for a particle moving in
a potential, but this analogy works only if allow the effective
temperature to vary with the position of the particle.

As with diffusion or Brownian motion, there is an alternative to the
diffusion equation for $P(n,t)$ and this is to write an equation
of motion for $n(t)$ which supplements   Eq. (\ref{kinetics}) by the
addition of a random or Langevin force
$\xi(t)$:
\begin{eqnarray}
{{dn}\over{dt}} &=& f(n) - g(n) + \xi (t),
\label{langevin}
\\
\langle \xi (t) \xi (t') \rangle &=& [f(n) + g(n)] \delta (t-t') .
\label{spectrum}
\end{eqnarray} 
From the Langevin equation we can also develop the distribution functional
for the probability of trajectories $n(t)$.  It should be emphasized
that all of these approaches are equivalent provided that we are
careful to treat the spatial variations of the effective
temperature \cite{zinn}.\footnote{In a review written for a biological
audience, McAdams and Arkin \cite{arkinTIG} state that Langevin methods
are unsound and can yield invalid predictions precisely for the case of
bistable reaction systems which interests us here; this is part of their
argument for the necessity of stochastic simulation methods as opposed to
analytic approaches.  Their reference for the failure of Langevin methods
\cite{badLangevin}, however, seems to consider only Langevin terms with
constant spectral density, thus ignoring (in the present language) the
spatial variations of effective temperature.  For the present problem
this would mean replacing   the noise correlation
function $[f(n) + g(n)]\delta(t-t')$ in  Eq. (\ref{spectrum}) by
$Q\delta(t-t')$ where $Q$ is a constant.  This indeed is wrong, and is
not equivalent to the master equation.  On the other hand, if the
arguments of Refs.
\cite{arkinTIG,badLangevin} were generally correct, they would imply
that Langevin methods could not used for the description of Brownian
motion with a spatially varying temperature, and this would be quite a
surprise.}  In one dimension this complication does not impede solving the
problem.  For any particular kinetic scheme we can compute the effective
potential and temperature, and kinetic schemes with multiple stable states
correspond to potential functions with multiple minima.

\section{Noise induced switching rates}

We want to know how the noise term destabilizes  the distinct stable
states of the switch.  If the noise is small, then by analogy with
thermal noise we expect that there will be some small jitter around the
stable states,  but also some rate of spontaneous jumping between the
states, analogous to thermal activation over an energy barrier as in
a chemical reaction.  This jumping rate
should be the product of an ``attempt frequency''---of order the
relaxation rate in the neighborhood of  one stable state---and a
``Boltzmann factor'' that expresses the exponentially small probability
of going over the barrier.  For ordinary chemical reactions this
Boltzmann factor is just $\exp(-F^\dag/k_B T)$, where $F^\dag$ is the 
activation free energy. If we want to build a switch that can be stable
for a time much longer than the switching time itself, then the Boltzmann
factor has to provide this large ratio of time scales.

There are several ways to calculate the analog of the Boltzmann factor for the
dynamics in Eq. (\ref{langevin}). 
The first step is to make more explicit the analogy with Brownian motion
and thermal activation. Recall that Brownian motion of an overdamped
particle is described by the Langevin equation
\begin{equation}
\gamma {{dx}\over {dt}} = -V' (x) + \eta (t) ,
\end{equation}
where $\gamma$ is drag coefficient of the particle, $V(x)$ is the potential, and
the noise force has correlations
$\langle \eta(t) \eta(t') \rangle = 2\gamma T\delta (t-t')$,
where $T$ is the absolute temperature measured in energy units so that
Boltzmann's constant is equal to one.  Comparing with Eq. (\ref{langevin}), we see that our
problem is equivalent to a particle with $\gamma = 1$  in an effective
potential $V_{\rm eff}(n)$ such that
$V_{\rm eff}'(n) = g(n) - f(n)$, at an effective temperature
$T_{\rm eff}(n) = [f(n) + g(n)]/2$.

If the temperature were uniform then the
equilibrium distribution of $n$ would be $P_{\rm eq}(n) \propto
\exp[-V_{\rm eff}(n)/T_{\rm eff}]$. With nonuniform temperature the
result is (up to weakly varying prefactors)
\begin{eqnarray}
P_{\rm eq}(n) &\propto& \exp[-U(n)]\\
U(n) &=& \int_0^n dy {{V_{\rm eff}'(y)}\over{T_{\rm eff}(y)}} .
\end{eqnarray}
One way to identify the Boltzmann factor for spontaneous switching is then to
compute the relative equilibrium occupancy of the stable states ($n_0$ and $n_1$)
and the unstable ``transition state'' at $n_*$.  The result is that the
effective activation energy for transitions from a stable state at $n =
n_0$ to the stable state at $n=n_1 > n_0$ is
\begin{equation}
F^\dag (n_0 \rightarrow n_1) = 2 k_B T\int_{n_0}^{n_*} dn
{{g(n) - f(n)}\over {g(n) + f(n)}} ,
\label{finalaction1}
\end{equation}
where $n_*$ is the unstable point,
and similarly for the reverse transition,
\begin{equation}
F^\dag (n_1 \rightarrow n_0) = 2 k_B T \int_{n_*}^{n_1} dn
{{f(n) - g(n)}\over {g(n) + f(n)}} .
\label{finalaction2}
\end{equation}

An alternative approach is to note that the
distribution of trajectories
$n(t)$ includes locally optimal paths that  carry the system from each
stable point up to the transition state; the effective activation free
energy can then be written as an integral along these optimal paths. 
The use of optimal path ideas in chemical
kinetics has a long history, going back at least to Onsager.  A
discussion in the spirit of the present one is Ref. \cite{dykman}.  For
equations of the general form
\begin{equation}
{{dn}\over{dt}} = -V_{\rm eff}'(n) + \xi(t),
\end{equation}
with $\langle \xi(t) \xi(t')\rangle = 2T_{\rm eff}(t)\delta(t-t')$,
the probability distribution for trajectories
$P[n(t)]$ can be written as \cite{zinn}
\begin{eqnarray}
P[n(t)] &=& \exp\left(-S[n(t)]\right)\\
S[n(t)] &=&
{1\over 4}\int dt {1\over{T_{\rm eff}(t)}}[{\dot n}(t) + V_{\rm eff}'(n(t))]^2
-{1\over 2} \int dt V_{\rm eff}''(n(t))  .
\label{actionA}
\end{eqnarray}
If the temperature $T_{\rm eff}$ is small, then the trajectories that minimize the action
should be determined primarily by minimizing the first  term in Eq. (\ref{actionA}),
which is $\sim 1/T_{\rm eff}$.  Identifying the effective potential and temperature as above,
the relevant term is
\begin{eqnarray}
{1\over 2} 
\int dt 
{{[{\dot n}  -f(n)+g(n)]^2}\over{f(n) + g(n)}}
&=& {1\over 2}\int dt {{{\dot
n}^2}\over{f(n) + g(n)}}
+ {1\over 2}\int dt {{[f(n) - g(n)]^2}\over{f(n)+g(n)}}
\nonumber\\
&&\,\,\,\,\,-\int dt {\dot n} {{f(n) - g(n)}\over{f(n) + g(n)}} .
\label{action}
\end{eqnarray}
We are searching for trajectories which take $n(t)$ from a stable point
$n_0$ where $f(n_0) = g(n_0)$ through the unstable point $n_*$ where $f$
and $g$ are again equal but the derivative of their difference (the
curvature of the potential) has changed sign.
For a discussion of the analogous quantum mechanical problem of
tunneling in a double well, see Ref. \cite{coleman}.
First we note that along any trajectory from $n_0$ to
$n_*$ we can simplify the third term in Eq. (\ref{action}):
\begin{equation}
\int dt {\dot n} {{f(n) - g(n)}\over{f(n) + g(n)}}
= \int_{n_0}^{n_*} dn  {{f(n) - g(n)}\over{f(n) + g(n)}} .
\end{equation}
This term thus depends on the endpoints of  the trajectory and not on the
path, and therefore cannot contribute to the structure of the optimal
path. In the analogy to mechanics, the first two terms are equivalent to
the (Euclidean) action for a particle with position dependent mass in a
potential; this means that along extremal trajectories there is a
conserved energy
\begin{equation}
E = {1\over 2} {{{\dot n}^2}\over{f(n) + g(n)}}
- {1\over 2}  {{[f(n) - g(n)]^2}\over{f(n)+g(n)}} .
\end{equation}
At the endpoints of the trajectory we have $\dot n  =0$ and $f(n) = g(n)$, and so we
are looking for zero energy trajectories, along which
\begin{equation}
{\dot n} (t) = \pm [f(n(t)) - g(n(t))] .
\end{equation}
Substituting back into Eq. (\ref{action}),  and being careful about the
signs, we find once again Eq's. (\ref{finalaction1},\ref{finalaction2}).

Both the `transition state' and the optimal path method involve
approximations, but if the noise is not too large the
approximations are good and the results of the two methods agree. 
Yet another approach is to solve the master equation (\ref{master})
directly, and again one gets the same answer for the switching rate when
the noise is small, as expected since  all the different approaches
are all equivalent if we make consistent approximations.  It is much more
work to find the prefactors of the rates, but we are concerned here with
orders of magnitude, and hence the prefactors aren't so important.

\section{Interpretation}

The crucial thing to notice in this calculation is that the integrands  in
Eq's. (\ref{finalaction1},\ref{finalaction2}) are bounded by one, so the
activation energy (in units of the thermal energy
$k_B T$) is bounded by twice the change in
the number of molecules.  
Translating back
to the spontaneous switching rates, the result is that the noise driven
switching time is longer than the relaxation time after switching by a
factor that is bounded,
\begin{equation}
{{\rm spontaneous\ switching\ time}
\over{\rm relaxation\ time}}
< \exp(\Delta n ),
\end{equation}
where $\Delta n$ is the change in the  number of molecules required to go
from one stable `switched' state to the other.  Imagine that we have a
reaction scheme in which the difference between the two stable states
corresponds to roughly 25 molecules.   Then it is possible to have  a
Boltzmann factor of up to $\exp({25}) \sim 10^{10}$.   
Usually we think of this as a limit to stability:  with
25 molecules we can have a Boltzmann  factor of no more than $\sim
10^{10}$.  But here I want to emphasize the positive statement that there
exist kinetic schemes in which just 25 molecules would be sufficient to
have this level of stability.  This corresponds to years per
millisecond:  with twenty five molecules, a biochemical switch that can
flip in milliseconds can be stable for years.  Real  chemical reaction
schemes will not saturate this bound, but certainly such stability is 
possible with roughly 100 molecules.  The genetic switch in
$\lambda$ phage operates with roughly 100 copies of the repressor
molecules \cite{ptashne}, and even in this simple system  there is
extreme stability:  the genetic switch is flipped spontaneously only once
in $10^5$ generations of the host bacterium
\cite{lambda-rev}.  Kinetic schemes with greater cooperativity get
closer to the bound, achieving greater stability for the same number of
molecules.

In electronics, the construction of digital
elements provides insulation against fluctuations on a microscopic scale
and allows a separation between the logical and physical design of a large
system.   We see that, once a cell has access to several tens of
molecules, it is possible to construct `digital' switch elements with
dynamics that are no longer significantly affected by microscopic
fluctuations.  Furthermore, weak interactions of these molecules with
other cellular components cannot change the basic `states' of the switch,
although these interactions can couple state changes to other events.

The importance of this `digitization'  on the scale of $10 -100$
molecules is illustrated by different models for pattern formation in
development.  In the classical model due to Turing, patterns are
expressed by spatial variations in the concentration of different
molecules, and patterns arise because uniform concentrations are rendered
unstable through the combination of nonlinearities in the kinetics with
the different diffusion constants of different substances. In this
picture, the spatial structure of the pattern is linked directly to
physical properties of the molecules.  An alternative  that each spatial
location is labelled by a set of discrete possible states, and patterns
evolve out of the `automaton' rules by which each location changes state
in relation to the neighboring states.  In this picture states and rules
are more abstract, and the dynamics of pattern formation is really at a
different level of description from the molecular dynamics of chemical
reactions and diffusion.    Reliable implementations of automaton rules
apparently are accessible as soon as the relevant chemical reactions
involve a few dozen molecules.

Biochemical switches have been reconstituted in vitro, but I am
not aware of any attempts to  verify that stable switching is possible
with small numbers of molecules.  It would be most interesting to study
model systems in which one could confine and monitor sufficiently few
molecules that it  becomes possible to observe spontaneous switching,
that is the breakdown of stability. Although genetic switches have
certain advantages,  even the simplest systems would require full
enzymatic apparatus for gene expression (but see Ref. \cite{albert} for
recent progress on controllable in vitro expression
systems).\footnote{Note also that reactions involving polymer synthesis
(mRNA from DNA or protein from mRNA) are not `elementary' reactions in the
sense described by Eq. (\ref{master}).  Synthesis of a single mRNA
molecule involves thousands of steps, each of which occurs (conditionally)
at constant probability per unit time, and so the noise in the overall
synthesis reaction is very different.  If the synthesis enzymes are
highly processive, so that the polymerization apparatus incoporates many
monomers into the polymer before `backing up' or falling off the
template, then synthesis itself involves a delay but relatively little
noise; the dominant source of noise becomes the assembly and disassembly
of the polymerization complex.  Thus there is some subtlety in trying to
relate a simple model to the complex sequence of reactions involved in
gene expression.  On the other hand a detailed simulation is problematic,
since there are so many different elementary steps with unknown rates. 
This combination of circumstances would make experiments on a minimal, in
vitro genetic switch espcially interesting.} Kinase switches are much
simpler, since they can be constructed from just a few proteins and can be
triggered by calcium; caged calcium  allows for an optical pulse to serve
as input.  

At reasonable protein concentrations,
$10-100$ molecules are found in a  volume of roughly 1 $(\mu{\rm m})^3$. 
Thus it should be possible to fabricate an array of `cells' with linear
dimensions ranging from 100 nm to 10 $\mu$m, such that solutions of 
kinase and accessory proteins would switch stably in the larger cells but
exhibit instability and spontaneous switching in the smaller cells.  The
state of the switch could be read out by including marker proteins that
would serve as substrates of the kinase but have, for example,
fluorescence lines that are shifted by phosphorylation, or by having
fluorescent probes on the kinase itself; transitions of single enzyme
molecules should be observable
\cite{1enzymea,1enzymeb}. 

A related idea would be to construct vesicles containing ligand gate ion
channels which can conduct calcium, and then have inside the vesicle
enzymes for synthesis and degradation of the ligand which are calcium
sensitive.  The cGMP channels of rod photoreceptors are an example, and
in rods the cyclase synthesizing cGMP is calcium sensitive, but the sign
is wrong to make a switch \cite{rods}; presumably this could solved by
appropriate mixing and matching of protein components from different
cells.  In such a vesicle the different stable states would be
distinguished by  different levels of internal calcium (as with
adaptation states in the rod), and these could be read out optically
using calcium indicators; caged calcium would again provide an optical
input to flip the switch. Amusingly, a close packed array of such
vesicles with
$\sim 100$ nm dimension would provide an optically addressable and
writable memory with storage density comparable to current RAM, albeit
with much slower switching.

In summary, it should be possible to build stable biochemical switches
from a few tens of molecules, and it seems likely that nature makes use of
these.  To test our understanding of stability we have to construct
systems which cross the threshold for observable instabilities, and this
seems accessible experimentally in several systems.

\subsubsection*{Acknowledgments}

Thanks to M. Dykman, J. J. Hopfield, and A. J. Libchaber for
helpful discussions.

\end{document}